\documentstyle[]{article}
\makeatletter
\def\emline#1#2#3#4#5#6{%
       \put(#1,#2){\special{em:moveto}}%
       \put(#4,#5){\special{em:lineto}}}
\def\newpic#1{}
\def\hybrid{\topmargin 0pt      \oddsidemargin 0pt
        \headheight 0pt \headsep 0pt
        \textwidth 6.25in       % A4 paper
        \textheight 9.5in       % A4 paper
        \marginparwidth 0.0in
        \parskip 5pt plus 1pt   \jot = 1.5ex}
\catcode`\@=11
\def\marginnote#1{}

\newcount\hour
\newcount\minute
\newtoks\amorpm
\hour=\time\divide\hour by60
\minute=\time{\multiply\hour by60 \global\advance\minute by-\hour}
\edef\standardtime{{\ifnum\hour<12 \global\amorpm={am}%
        \else\global\amorpm={pm}\advance\hour by-12 \fi
        \ifnum\hour=0 \hour=12 \fi
        \number\hour:\ifnum\minute<10 0\fi\number\minute\the\amorpm}}
\edef\militarytime{\number\hour:\ifnum\minute<10 0\fi\number\minute}

\def\draftlabel#1{{\@bsphack\if@filesw {\let\thepage\relax
   \xdef\@gtempa{\write\@auxout{\string
      \newlabel{#1}{{\@currentlabel}{\thepage}}}}}\@gtempa
   \if@nobreak \ifvmode\nobreak\fi\fi\fi\@esphack}
        \gdef\@eqnlabel{#1}}
\def\@eqnlabel{}
\def\@vacuum{}
\def\draftmarginnote#1{\marginpar{\raggedright\scriptsize\tt#1}}

\def\draftlabel#1{{\@bsphack\if@filesw {\let\thepage\relax
   \xdef\@gtempa{\write\@auxout{\string
      \newlabel{#1}{{\@currentlabel}{\thepage}}}}}\@gtempa
   \if@nobreak \ifvmode\nobreak\fi\fi\fi\@esphack}
        \gdef\@eqnlabel{#1}}
\def\@eqnlabel{}
\def\@vacuum{}
\def\draftmarginnote#1{\marginpar{\raggedright\scriptsize\tt#1}}

\def\draft{\oddsidemargin -.5truein
        \def\@oddfoot{\sl preliminary draft \hfil
        \rm\thepage\hfil\sl\today\quad\militarytime}
        \let\@evenfoot\@oddfoot \overfullrule 3pt
        \let\label=\draftlabel
        \let\marginnote=\draftmarginnote
   \def\@eqnnum{(\theequation)\rlap{\kern\marginparsep\tt\@eqnlabel}%
\global\let\@eqnlabel\@vacuum}  }

\def\numberbysection{\@addtoreset{equation}{section}
        \def\theequation{\thesection.\arabic{equation}}}

\def\underline#1{\relax\ifmmode\@@underline#1\else
        $\@@underline{\hbox{#1}}$\relax\fi}

\def\titlepage{\@restonecolfalse\if@twocolumn\@restonecoltrue\onecolumn
     \else \newpage \fi \thispagestyle{empty}\c@page\z@
        \def\thefootnote{\fnsymbol{footnote}} }

\def\endtitlepage{\if@restonecol\twocolumn \else  \fi
        \def\thefootnote{\arabic{footnote}}
        \setcounter{footnote}{0}}  %\c@footnote\z@ }
\relax

\draft

\numberbysection
\hybrid

\def\beq{\begin{equation}}
\def\eeq{\end{equation}}

\newdimen\Squaresize \Squaresize=30pt
\newdimen\Thickness \Thickness=0.5pt

\def\Square#1{\hbox{\vrule width \Thickness
   \vbox to \Squaresize{\hrule height \Thickness\vss
      \hbox to \Squaresize{\hss#1\hss}
   \vss\hrule height\Thickness}
\unskip\vrule width \Thickness}    %Example:
\kern-\Thickness}                  %\young{1&2&3&4\cr 5&6&7\cr 8&9\cr 10\cr}

\def\Vsquare#1{\vbox{\Square{$#1$}}\kern-\Thickness}

\begin{document}

\begin{titlepage}

\title{Zero curvature representation for classical lattice
sine-Gordon equation via quantum $R$-matrix}

\author{A.Zabrodin\thanks{Joint Institute of
Chemical Physics, Kosygina str. 4, 117334,
Moscow, Russia and ITEP, 117259, Moscow, Russia}}
\date{September 1997}
\maketitle

\begin{abstract}

Local $M$-operators for the classical sine-Gordon model in
discrete space-time are constructed by convolution of the
quantum trigonometric 4$\times$4 $R$-matrix with certain vectors
in its "quantum" space. Components of the vectors are identified
with $\tau$-functions of the model. This construction
generalizes the known representation of $M$-operators
in continuous time models in terms of Lax operators
and classical $r$-matrix.

\end{abstract}

\vfill

\end{titlepage}

\section{Introduction}

The $r$-matrix approach allows one to develop a unified
treatment \cite{FadTakh} of
non-linear integrable equations as
hamiltonian systems having enough number of conserved
quantities in involution.
The role of the (classical) $r$-matrix is to provide a universal
form of Poisson brackets for elements of Lax operators.

An alternative approach \cite{FadTakh},\,\cite{ZMNP}
consists in representing
non-linear equations as 2D zero curvature conditions
for a pair of matrix functions (called $L$ and $M$ operators)
depending on a spectral parameter.
Although this method avoids any reference to the hamiltonian
aspects of the problem, the $r$-matrix is meaningful here, too.
The alternative though less popular point of view
on the $r$-matrix (which we are going to follow)
is to treat it as a machine to produce $M$-operators from
$L$-operators. Let us recall how it works.

Let ${\cal L}_l (z)$ be a classical
2$\times$2 $L$-operator on 1D lattice with the periodic
boundary condition ${\cal L}_{l+N}(z)={\cal L}_{l}(z)$;
$z$ is the spectral parameter.
We assume the ultralocality -- Poisson
brackets of dynamical variables at different sites are
equal to zero.
The monodromy matrix ${\cal T}_{l}(z)$ is ordered product
of $L$-operators along the lattice from the site $l$ to
$l+N-1$:
\beq
{\cal T}_{l}(z) =\prod _{l+N-1\geq j \geq l}^{\longleftarrow}
{\cal L}_{j}(z)\,.
\label{r1}
\eeq
Generating function of
conserved quantities
is obtained by taking trace of the monodromy matrix:
\beq
T(z)=\mbox{tr}\, {\cal T}_{l}(z)
\label{r2}
\eeq
which does not depend on $l$ due to the periodic boundary
condition. Conserved quantities obtained by
expanding $\mbox{log}T(z)$ in $z$ are hamiltonians of
commuting flows.

All these flows admit a zero curvature representation.
The generating function of corresponding
$M$-operators is given by \cite{Sklyanin},\,\cite{FadTakh}
\beq
M_{l}(z;w) =T^{-1}(w)\, \mbox{tr}_{1}
\left [ r\big (\frac{z}{w}\big )
({\cal T}_{l}(w)\otimes I) \right ],
\label{r3}
\eeq
where $I$ is the unity matrix,
$r(z)$ is the (classical) 4$\times$4 $r$-matrix.
It acts in the tensor product of two 2-dimensional
spaces and, therefore, has natural block structure.
In (\ref{r3}), $\mbox{tr}_{1}$ means trace in the first space.
Expanding the r.h.s. of eq.\,(\ref{r3})
in $w$, one gets $M$-operators depending on the spectral
parameter $z$.
From the hamiltonian point of view,
the zero curvature condition follows from $r$-matrix Poisson
brackets for elements of the $L$-operator.
In general, $M_{l}(z;w)$
is a non-local quantity.

A way to construct local $M$-operators from the generating
function is well known \cite{IK},\,\cite{FadTakh},\,\cite{book}.
Suppose there exists a value $z_0$ of the spectral
parameter such that the $L$-operator degenerates for all $l$,
i.e. $ \det {\cal L}_l (z_0 )=0$. This means that
${\cal L}_{l}(z_0 )$ is
a 1-dimensional projector:
\beq
{\cal L}_l (z_0 )=\frac{\big | \alpha _{l}\big >
\big < \beta _{l} \big |}{\lambda _{l}}\,,
\label{r4}
\eeq
where
\beq
\big | \alpha \big >=
\left (\! \begin{array}{c} \alpha ^{(1)}\\ \alpha ^{(2)}\end{array}
\! \right ),
\;\;\;\;\;\;\;\;
\big < \beta \big |=\big ( \beta ^{(1)},\, \beta ^{(2)}\big )
\label{r5}
\eeq
are two-component vectors and $\lambda _{l}$ is a scalar
normalization factor. (Of course this factor can be included
in the vectors.)
Components of the vectors as well as the $\lambda _{l}$
depend on dynamical variables.
It is easy to see that $M_{l}(z;z_0 )$ is a local quantity:
\beq
M_l (z)\equiv  M_{l}(z; z_0 )=
\frac{ \big < \beta _{l}
\big | r(z/z_0 )\big |\alpha _{l-1} \big >}
{\big < \beta _{l}\big | \alpha _{l-1}\big >}
\label{r6}
\eeq
(note that the normalization factor cancels).
The scalar product is taken in the first space only, so the result
is a 2$\times$2 matrix with the spectral parameter $z$.
It generates infinitezimal
continuous time shifts. The zero curvature condition
giving rise to non-linear integrable equations reads
\beq
\partial _{t}{\cal L}_{l}(z)=
M_{l+1}(z){\cal L}_{l}(z)-
{\cal L}_{l}(z)M_{l}(z)\,.
\label{r7}
\eeq

The goal of this work is to
extend eq.\,(\ref{r6}) to fully discretized analogues of classical
2D integrable models and to obtain a similar representation
for $M$-operators corresponding to {\it discrete time flows}.
By the fully discretized models we mean a family of 2D
partial difference integrable equations introduced and
studied by Hirota \cite{HirotaKdV}-\cite{Hirota}:
discrete analogues of the KdV,
the Toda chain, the sine-Gordon (SG) equations, etc.
Our guiding principle is Miwa's interpretation \cite{Miwa} of
the discrete time flows, in which discretized
integrable equations are treated as members
of the same infinite hierarchy as their continuous counterparts.
(This idea was further developed in \cite{DJM} as a general method
to produce discrete soliton equations from continuous ones.)
Although in this paper we deal with the discrete SG equation only,
there is no doubt that the results can be more or less
straightforwardly extended to other integrable models on 2D lattice.

Let us outline the results. A formula for discrete $M$-operators
of the type (\ref{r6}) does exist.
Remarkably, the structure of the formula remains the same, with
the classical $r$-matrix
being substituted by quantum $R$-matrix.
Specifically, we show that the following representation of discrete
$M$-operators holds:
\beq
{\cal M}_l (z)=
\frac{ \big < \beta _{l}
\big | R(z/z_0 )\big |\check \beta _{l-1} \big >}
{\big < \beta _{l}\big | \alpha _{l-1}\big >}\,,
\;\;\;\;\;\;\;\;
\big |\check \beta _{l} \big >\equiv
\sigma _{1}\big | \beta _{l} \big >
\label{r8}
\eeq
(here and below $\sigma _{1}, \sigma _{2}, \sigma _{3}$ are
the Pauli matrices). In the r.h.s.,
$R(z)$ is a {\it quantum} 4$\times$4 $R$-matrix to be
specified below with the "quantum" parameter $q$ related to the
discrete time lattice spacing. We stress that vectors
$\big | \alpha _{l} \big >$ and
$\big | \beta _{l} \big >$ in this formula are
{\it the same} as in eq.\,(\ref{r6}).

The $M$-operator (\ref{r8}) generates shifts of a discrete
time variable $m$.
The discrete zero curvature condition
\beq
{\cal M}_{l+1,m}(z){\cal L}_{l,m}(z)
={\cal L}_{l,m+1}(z){\cal M}_{l,m}(z)
\label{dzc}
\eeq
(with the same $L$-operator as before) gives rise to
non-linear integrable equations in discrete space-time
listed above.
Under the condition that these equations
are satisfied
a similar $R$-matrix representation for the
$L$-operator itself is valid:
\beq
{\cal L}_l (z)=
\frac{ \big < \beta _{l}
\big | R^{(-)}(z/z_0 )\big |\alpha _{l} \big >}
{\big < \beta _{l}\big | \alpha _{l-1}\big >}\,.
\label{r9}
\eeq
Here $R^{(-)}(z)$ is another quantum $R$-matrix which
differs from $R(z)$ by Drinfeld's twist.

As soon as the quantum $R$-matrix comes into play, we can
use customary conventions of the
algebraic Bethe ansatz approach \cite{FT}\,,\cite{book}
to represent formulas (\ref{r8}), (\ref{r9}) in the figure:

\begin{center}
\special{em:linewidth 0.4pt}
\unitlength 0.7mm
\linethickness{0.4pt}
\begin{picture}(90.00,59.00)(25.00,10.00)
\emline{60.00}{40.00}{1}{90.00}{40.00}{2}
\emline{75.00}{55.00}{3}{75.00}{25.00}{4}
\put(75.00,59.00){\makebox(0,0)[cc]{$\big <\beta \big |$}}
\put(75.00,20.00){\makebox(0,0)[cc]{$\big |\alpha \big >$}}
\put(52.00,40.00){\makebox(0,0)[rc]
{$\big < \beta \big | R(z)\big |\alpha \big >\, =\, $}}
\end{picture}
\end{center}

\vspace{-0.5cm}

\noindent
The scalar product is taken in the "quantum" (vertical) space
of the $R$-matrix, so the result is a $2\times 2$ matrix in
its auxiliary (horizontal) space.

The change of dynamical variables to the pair of vectors
$\big | \alpha _{l} \big >$,
$\big | \beta _{l} \big >$
plays a key role in the construction. In the papers \cite{IK}
on exact lattice regularization of integrable models
components of these vectors were expresed in terms of
canonical variables of the model. Those formulas looked
complicated and were hardly considered as something
instructive. Here we add a remark which may help to
understand their meaning.
Taking into account equations of motion of the completely
discretized model derived from the discrete zero curvature
condition (\ref{dzc}), we show that (suitably normalized)
components of the vectors
$\big | \alpha _{l} \big >$,
$\big | \beta _{l} \big >$
are {\it $\tau$-functions}. The $\tau$-function is known to be
one of the most fundamental objects of the theory
(see e.g. \cite{tau}).

In this paper we explain the construction for the simplest
example -- the lattice SG model.
In the literature, there are two lattice versions of the
classical SG model:
Hirota's discrete analogue of the SG
equation on a space-time lattice \cite{Hirota}
and the model on a space lattice with continuous time
introduced by Izergin and Korepin \cite{IK} and further
studied in \cite{Tarasov}.
Actually, they are closely connected with each other
and, furthermore, each of the two models can be better understood
with the help of the other one. They have common $L$-operator
(which can be represented in the form (\ref{r9})).
The $M$-operators are different: in the latter case
the r.h.s. of eq.\,(\ref{r6})
with the classical trigonometric $r$-matrix generates continuous time
derivatives while in the former case
discrete time shifts are generated by
(\ref{r8}). The $R$-matrix
entering the r.h.s. of this formula is in this
case the simplest trigonometric solution of the
Yang-Baxter equation (the $R$-matrix of the $XXZ$ spin chain).
At last, we demonstrate how eq.\,(\ref{r6}) is reproduced
from eq.\,(\ref{r8}) in the continuous time limit.

\section{Lattice SG models with discrete and continuous
time}

By the discrete SG model on the space-time lattice we
mean the Faddeev-Volkov version \cite{FadVol} of Hirota's
discrete analogue \cite{Hirota} of the classical SG equation:
\begin{eqnarray}
&&\nu \psi (u,v+1)\psi (u+1, v+1)-\nu
\psi (u,v)\psi (u+1, v)
\nonumber \\
&=&\mu \psi (u+1,v)\psi (u+1, v+1)-\mu
\psi (u,v)\psi (u, v+1)\,.
\label{nl1}
\end{eqnarray}
This is a non-linear partial difference equation
for a function $\psi$ of the two variables $u,v$, where
$\mu , \nu $ are constant parameters.
We call $u,v$ {\it chiral space-time coordinates}.
It was shown \cite{FadVol}, \cite{Volkov}
that eq.\,(\ref{nl1}) contains both SG and KdV
equations as results of different continuum limits.

Let

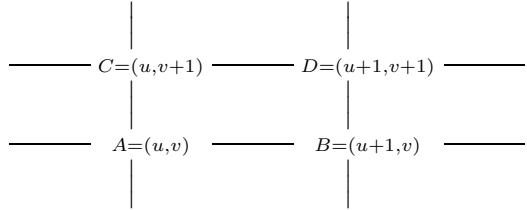
\vspace{0.1cm}

$$
\begin{array}{ccccc}
&\left |\phantom{\frac{A^A}{A^A}}\right. & &
\left |\phantom{\frac{A^A}{A^A}}\right. & \\
\frac{\phantom{aaaaaaa}}{\phantom{aaaaaaa}}\!\!\!&
\!\!\scriptstyle{C=(u,v+1)}\!\!&
\!\!\!\frac{\phantom{aaaaaaa}}{\phantom{aaaaaaa}}\!\!\!
&\!\!\scriptstyle{D=(u+1, v+1)}\!\!&
\!\!\!\frac{\phantom{aaaaaaa}}{\phantom{aaaaaaa}}\\
&\left |\phantom{\frac{A^A}{A^A}}\right. & &
\left |\phantom{\frac{A^A}{A^A}}\right. & \\
\frac{\phantom{aaaaaaa}}{\phantom{aaaaaaa}}\!\!\!&
\!\!\scriptstyle{A=(u,v)}\!\!&
\!\!\!\frac{\phantom{aaaaaaa}}{\phantom{aaaaaaa}}\!\!\!
&\!\!\scriptstyle{B=(u+1, v)}\!\!&
\!\!\!\frac{\phantom{aaaaaaa}}{\phantom{aaaaaaa}}\\
&\left |\phantom{\frac{A^A}{A^A}}\right. & &
\left |\phantom{\frac{A^A}{A^A}}\right. & \\
\end{array}
$$

\vspace{0.2cm}

\noindent
be an elementary cell of the $u,v$-lattice;
in this notation eq.\,(\ref{nl1}) reads
\beq
\psi _{C}\psi _{D}-\psi _{A}\psi _{B}
=\frac{\mu}{\nu}\big (\psi _{B}\psi _{D}-
\psi _{A}\psi _{C}\big )\,.
\label{nl3}
\eeq

The non-linear equation (\ref{nl3}) can be
represented \cite{FadVol} as a
discrete zero curvature condition. Set
\beq
\phi (u,v)\equiv \psi ^{\frac{1}{2}}(u,v)
\label{zc1}
\eeq
and consider the chiral $L$-operator \cite{FadVol},\,\cite{Fad}
\beq
L_{B\leftarrow A}(z;\mu )=
\left ( \begin{array}{ccc}
\mu \phi _{B}\phi _{A}^{-1} &&
z \phi _{B}^{-1}\phi _{A}^{-1}\\
&& \\
z \phi _{B}\phi _{A}&&
\mu  \phi _{B}^{-1}\phi _{A}\end{array} \right ).
\label{zc2}
\eeq
Equation (\ref{nl3}) is equivalent to the
zero curvature condition
\beq
L_{D\leftarrow B}(z;\nu )
L_{B\leftarrow A}(z;\mu )=
L_{D\leftarrow C}(z;\mu )
L_{C\leftarrow A}(z;\nu )
\label{zc3}
\eeq
on the chiral space-time lattice.

The chiral $L$-operators are building blocks for
more complicated ones.
For our purpose we need to pass from chiral to the "direct"
space-time coordinates $l=\frac{1}{2}(u+v)$,
$m=\frac{1}{2}(u-v)$. (From now on we refer to $l$ (resp., $m$)
as discrete space (resp., discrete time) coordinate.)
To do that, consider "composite" operators
which generate shifts along
the diagonals $A\rightarrow D$ and $C\rightarrow B$
respectively:
\beq
\hat {\cal L}_{D\leftarrow A}(z)=z^{-1}
L_{D\leftarrow C}(z;\mu )
L_{C\leftarrow A}(z;\nu ),
\label{c1}
\eeq
\beq
\hat {\cal M}_{B\leftarrow C}(z)= z^{-1}
(z^2 -\nu ^{2}) L_{B\leftarrow A}(z;\mu )
\left [L_{C\leftarrow A}(z;\nu )\right ]^{-1}.
\label{c3}
\eeq
Using (\ref{zc2}), we have:

\beq
\hat {\cal L}_{D\leftarrow A}(\mu z)=
\left ( \begin{array}{ccc}
\mu  \displaystyle{\frac{\phi _A}{\phi _D}}z
+\nu \displaystyle{\frac{\phi _D}{\phi _A}}z^{-1}
&&
\phi _{C}^{-2}\left (\mu \displaystyle{\frac{\phi _D}{\phi _A}}
+\nu \displaystyle{\frac{\phi _A}{\phi _D}}\right )\\ &&\\
\phi _{C}^{2}\left (\mu \displaystyle{\frac{\phi _A}{\phi _D}}
+\nu \displaystyle{\frac{\phi _D}{\phi _A}}\right )
&&
\mu \displaystyle{\frac{\phi _D}{\phi _A}}z
+\nu \displaystyle{\frac{\phi _A}{\phi _D}}z^{-1}
\end{array}\right ),
\label{c2}
\eeq

\beq
\hat {\cal M}_{B\leftarrow C}(\mu z)=
\left ( \begin{array}{ccc}
\mu  \displaystyle{\frac{\phi _C}{\phi _B}}z
-\nu \displaystyle{\frac{\phi _B}{\phi _C}}z^{-1}
&&
\phi _{A}^{-2}\left (\mu \displaystyle{\frac{\phi _B}{\phi _C}}
-\nu \displaystyle{\frac{\phi _C}{\phi _B}}\right )\\ &&\\
\phi _{A}^{2}\left (\mu \displaystyle{\frac{\phi _C}{\phi _B}}
-\nu \displaystyle{\frac{\phi _B}{\phi _C}}\right )
&&
\mu \displaystyle{\frac{\phi _B}{\phi _C}}z
-\nu \displaystyle{\frac{\phi _C}{\phi _B}}z^{-1}
\end{array}\right ).
\label{c4}
\eeq

Discrete zero curvature condition for these operators
has the form (\ref{dzc}). It gives rise to a system of non-linear
equations for fields at the vertices of
four neighbouring elementary cells, which
is a direct corollary of the basic equation (\ref{nl1}).

Let us turn to the lattice SG model with continuous time
introduced by Izergin and Korepin in \cite{IK}.
Its $L$-operator on $l$-th site has
the form\footnote{We use a slightly modified form
of the $L$-operator originally suggested in \cite{IK}.
They differ by
a unitary transformation and multiplication
by the matrix $\sigma _{2}$ from the left. The latter
modification takes into account the fact that we deal
with the Faddeev-Volkov equation (\ref{nl1}) rather than
Hirota's discrete SG equation itself.}
\beq
\hat {\cal L}_{l}^{(IK)}(z)=
\left ( \begin{array}{ccc}
z\chi _{l} +z^{-1}\chi _{l}^{-1}&&
s^{-\frac{1}{2}}\varphi _{l}\pi _{l} \\&&\\
s^{-\frac{1}{2}}\varphi _{l}\pi _{l}^{-1}&&
z\chi _{l}^{-1}+z^{-1}\chi _{l} \end{array} \right ).
\label{c5}
\eeq
Here $\pi _{l}$, $\chi _{l}$ are exponentiated canonical
variables on the lattice,
$\varphi _{l}=\left [ 1+s(\chi _{l}^{2}+\chi _{l}^{-2})
\right ]^{\frac{1}{2}}$
and $s$ is a parameter of the model.

As is known,
discretizing time in integrable models consists merely in
changing the $M$-operator and leaving the $L$-operator
unchanged, so we are to identify the $L$-operators (\ref{c5})
and (\ref{c2}). For that purpose, consider composite
fields
\beq
\pi (u,v)=\left [ \psi (u+1, v)\psi (u,v+1)\right ]^{\frac{1}{2}},
\;\;\;\;\;\;\;\;
\chi (u,v)=\left [ \frac{\psi (u,v)}{\psi (u+1, v+1)}
\right ] ^{\frac{1}{2}}
\label{c6}
\eeq
on the chiral lattice and set $\pi _{l}=\pi (l,l)$,
$\chi _{l}= \chi (l,l)$ at the constant time slice $m=0$.
Finally, identifying $s=\mu \nu (\mu^{2}+\nu ^{2})^{-1}$
and taking into account the equation of motion (\ref{nl1}),
we conclude that
\beq
\hat {\cal L}_{l}^{(IK)}(z)=(\mu \nu )^{-\frac{1}{2}}
\hat {\cal L}_{l}\big ((\mu \nu )^{\frac{1}{2}}z\big )\,.
\label{c7}
\eeq
Here we use the natural notation
\beq
\hat {\cal L}_{l}(z)\equiv
\hat {\cal L}_{D_{l}\leftarrow A_{l}}(z)\,,
\label{nn}
\eeq
where $A_l =(l,l),\,
D_l =(l+1, l+1)$ (note that in this notation
$D_{l}=A_{l+1}$). (In what follows we use the similar notation
for the $M$-operator (\ref{c4}):
$\hat {\cal M}_{\bar B_{l}\leftarrow A_{l}}(z)
\equiv \hat {\cal M}_{l}(z)$, where
$\bar B_{l} = (l+1, l-1)$.)

The $L$-operator $\hat {\cal L}_{l}^{(IK)}(z)$
has two degeneracy points
$z_{0}^{\pm }$. In terms of the parameters $\mu , \nu$
they are $z_{0}^{\pm}=(\mu /\nu )^{\pm \frac{1}{2}}$.
Using (\ref{c7}) and (\ref{c2}), it is not difficult to
represent $\hat {\cal L}_{l}^{(IK)}(z_{0}^{\pm})$
in the form (\ref{r4}) with the r.h.s. expressed through
the field $\psi (u,v)$.

\section{Hirota's bilinear formalism}

The idea of Hirota's approach is to treat eq.\,(\ref{nl1}) as a
consequence of 3-term bilinear equations
for $\tau$-functions, which are simpler and
in a sense more fundamental.
In this section we give a minimal extraction from
Hirota's method necessary for what follows (for more
details see e.g. \cite{DJM},\,\cite{Zab}).

In the case at hand we
need two $\tau$-functions: $\tau$ and $\hat \tau$.
Making the substitution
\beq
\psi (u,v)=\frac{\hat \tau (u,v)}{\tau (u,v)}\,,
\label{b1}
\eeq
we immediately see that eq.\,(\ref{nl1}) follows from
the couple of Hirota's bilinear equations \cite{Hirota}
\begin{eqnarray}
&&(\nu - \mu )\hat \tau _{A}\tau _{D}=
\nu \tau _{B}\hat \tau _{C}
-\mu \hat \tau _{B} \tau _{C}\,,
\nonumber \\
&&(\nu - \mu )\tau _{A}\hat \tau _{D}=
\nu \hat \tau _{B}\tau _{C}
-\mu \tau _{B} \hat \tau _{C}\,.
\label{b2}
\end{eqnarray}
The equivalent form  of these equations,
\begin{eqnarray}
&&(\nu + \mu ) \tau _{B}\hat \tau _{C}=
\mu \tau _{A}\hat \tau _{D}
+\nu \hat \tau _{A} \tau _{D}\,,
\nonumber \\
&&(\nu + \mu )\hat \tau _{B} \tau _{C}=
\mu \hat \tau _{A}\tau _{D}
+\nu \tau _{A} \hat \tau _{D}\,,
\label{b3}
\end{eqnarray}
is equally useful in what follows.
At last, we point out the relation
\beq
\tau (u-1, v)\hat \tau (u+1, v)
+\hat \tau (u-1, v)\tau (u+1, v)
=2\tau (u, v)\hat \tau (u, v)
\label{b4}
\eeq
valid for the $\tau$-functions of the discrete SG model.

Let us give some additional remarks.
Equations (\ref{b2}) are to be thought of as being embedded into
the infinite 2D Toda lattice hierarchy with 2-periodic
reduction \cite{UT}. This embedding implies that
parameters $\mu$ and $\nu$ are Miwa's variables \cite{Miwa}
corresponding to the two elementary discrete flows $u,v$.
Miwa's variables play the role of inverse lattice spacings
in the chiral directions. Lattice spacing in the $m$-direction
is then to be identified with $\frac{\mu -\nu}{\mu \nu}$.
One should note, however, that
in this approach the chiral coordinate axes
are in general not orthogonal
to each other. In particular,
as it is seen from eqs.\,(\ref{b2}), at $\mu =\nu$ one
must {\it identify} $u$ and $v$, so that the 2D lattice collapses
to a 1D one. In this sense eq.\,(\ref{b4}) follows from
(\ref{b3}) at $\nu = \mu$.

\section{Quantum $R$-matrix in the classical lattice SG model}

Our goal is to represent the $L$ and $M$ operators
(\ref{c2}), (\ref{c4}) as convolutions of quantum $R$-matrix
with some vectors in the "quantum" space.

Consider quantum $R$-matrices of the form
\beq
R^{(\pm )}(z)=\left (
\begin{array}{ccccccc}
a(z)&&0&&0&&0 \\
&&&&&&\\
0&&\pm b(z) && c(z) && 0 \\
&&&&&&\\
0 && c(z) &&\pm b(z) && 0 \\
&&&&&&\\
0&& 0 && 0&&  a(z)
\end{array} \right ),
\label{q1}
\eeq
where
\beq
a(z)=qz-q^{-1}z^{-1}, \;\;\;\;
b(z)=z-z^{-1}, \;\;\;\;
c(z)=q-q^{-1}.
\label{q2}
\eeq
Here $q$ is a "quantum" parameter and $z$ is the spectral parameter.
The $R$-matrices $R^{(+)}$ and $R^{(-})$
differ by Drinfeld's twist.
Both of them satisfy the quantum Yang-Baxter equation.
In the Introduction we have used the notation
$R^{(+)}(z)=R(z)$. When necessary, we write
$R(z)=R(z;q)$.

The $R$-matrix has 2$\times$2 block structure with respect
to the "auxiliary" space. Let $i, i'$ number block rows and
columns and let $j, j'$ number rows and columns inside each block
(i.e., in the "quantum" space). Then matrix elements of the
$R$-matrix
of the form (\ref{q1}) are denoted by $R(z)^{ii'}_{jj'}$.
Let $\big |\alpha \big >$,
$\big |\beta \big >$
be two vectors in the quantum space (see (\ref{r5})).
Each block of the
$R$-matrix is an operator in the quantum space. Consider
its action to $\big |\alpha \big >$ and subsequent scalar product with
$\big <\beta \big |$ (see the figure in the Introduction).
The result
is a 2$\times$2 matrix in the auxiliary space:
$$
\big <\beta \big | R(z) \big |\alpha \big >_{ii'}
=\sum _{jj'}R(z)^{ii'}_{jj'}\alpha ^{(j')}\beta ^{(j)}.
$$
Substituting the matrices (\ref{q1}), we find:
\beq
\big <\beta \big | R^{(\pm )}(z) \big |\alpha \big > =\left (
\begin{array}{ccc}
\beta ^{(1)}\alpha ^{(1)}a(z) \pm \beta ^{(2)}\alpha ^{(2)}b(z) &&
\beta ^{(2)}\alpha ^{(1)}c(z) \\ && \\
\beta ^{(1)}\alpha ^{(2)}c(z) &&
\pm \beta ^{(1)}\alpha ^{(1)}b(z) +\beta ^{(2)}\alpha ^{(2)}a(z)
\end{array} \right ).
\label{q3}
\eeq

We are going to identify r.h.s. of eqs.\,(\ref{c2}),
(\ref{c4}) with this matrix. The key step is to
express matrix
elements of the $L$ and $M$ operators through the
$\tau$-functions according to (\ref{b1}) and after that
make use of Hirota's bilinear equations (\ref{b2}), (\ref{b3})
when necessary.
The best result is achieved after
the diagonal gauge transformation
\beq
\hat {\cal L}_{D \leftarrow A}(z)
\, \longrightarrow \,
{\cal L}_{A\leftarrow D}(z)=
\left ( \frac{\tau _{D}\hat \tau _{D}}
{\tau _{A}\hat \tau _{A}} \right )^{\frac{1}{2}}
\hat {\cal L}_{D\leftarrow A}(z)\,,
\label{q4a}
\eeq
\beq
\hat {\cal M}_{B\leftarrow C}(z)
\, \longrightarrow \,
{\cal M}_{B\leftarrow C}(z)=
\left ( \frac{\tau _{B}\hat \tau _{B}}
{\tau _{C}\hat \tau _{C}} \right )^{\frac{1}{2}}
\hat {\cal M}_{B\leftarrow C}(z)\,.
\label{q4b}
\eeq

Omitting details of the computation, we present
the final result. Set
\beq
\big |\alpha \big >=
\left ( \begin{array}{c} \tau  \\
\hat \tau \end{array} \right ), \;\;\;\;\;\;\;\;
\big |\beta \big > =
\left ( \begin{array}{c} \hat \tau \\
\tau \end{array} \right ),
\;\;\;\;\;\;\;\;
q=\frac{\mu}{\nu}\,.
\label{q5}
\eeq
Using eqs.\,(\ref{b2}), (\ref{b3}), we get:
\beq
\big <\beta _{B}\big | R^{(-)}(z) \big |\alpha _{C}\big >
=\frac{\mu -\nu}{\mu \nu}
\tau _{A}\hat \tau _{A}
{\cal L}_{D\leftarrow A}(\mu z)\,,
\label{q7}
\eeq
\beq
\big <\beta _{D}\big | R^{(+)}(z) \big |\alpha _{A}\big >
=\frac{\mu +\nu}{\mu \nu}
\tau _{C}\hat \tau _{C}
{\cal M}_{B\leftarrow C}(\mu z)\,.
\label{q8}
\eeq
Finally, let us pass to the discrete space-time coordinates $l$, $m$
and consider the slice of the lattice at $m=0$.
The notations of the type (\ref{nn})
explained in Sect.2 are again convenient.
Making use of eq.\,(\ref{b4}), we arrive at
the formulas

\beq
{\cal L}_{l}(\mu z)=\frac{2\mu \nu}{\mu -\nu}\,
\frac{
\big <\beta _{l}\big | R^{(-)}(z) \big |\alpha _{l}\big >}
{\big <\beta _{l} \big |\alpha _{l-1}\big >}\,,
\label{q9}
\eeq

\beq
{\cal M}_{l}(\mu z)=\frac{2\mu \nu}{\mu +\nu}\,
\frac{
\big <\beta _{l}\big | R^{(+)}(z) \big |\check \beta _{l-1}\big >}
{\big <\beta _{l} \big |\alpha _{l-1}\big >}\,,
\label{q10}
\eeq

\noindent
which up to the constant prefactors coincide with the ones
announced in the Introduction.
Location of the vectors
$|\alpha _{l}\big >$, $|\beta _{l}\big >$
on the lattice with respect to the $m=0$ slice
is shown in the figure:

\vspace{0.2cm}

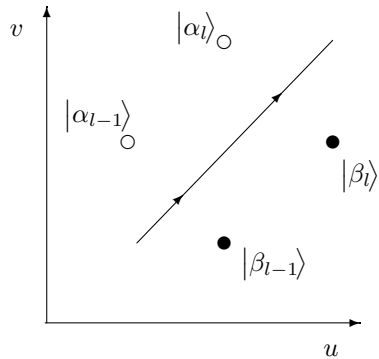
\begin{center}
\special{em:linewidth 0.4pt}
\unitlength 0.6mm
\linethickness{0.4pt}
\begin{picture}(84.00,80.00)
\put(80.00,10.00){\vector(1,0){0.2}}
\emline{10.33}{10.00}{1}{80.00}{10.00}{2}
\put(10.67,80.00){\vector(0,1){0.2}}
\emline{10.67}{10.00}{3}{10.67}{80.00}{4}
\emline{74.00}{72.67}{5}{30.67}{27.67}{6}
\put(50.00,27.67){\circle*{2.67}}
\put(74.00,50.00){\circle*{2.67}}
\put(50.00,72.33){\circle{2.67}}
\put(28.67,50.00){\circle{2.67}}
\put(4.00,75.00){\makebox(0,0)[cc]{$v$}}
\put(73.67,4.00){\makebox(0,0)[cc]{$u$}}
\put(61.00,23.00){\makebox(0,0)[cc]{$\big |\beta _{l-1}\!\big >$}}
\put(80.00,42.67){\makebox(0,0)[cc]{$\big |\beta _{l}\!\big >$}}
\put(22.00,55.90){\makebox(0,0)[cc]{$\big |\alpha _{l-1}\!\big >$}}
\put(43.67,75.33){\makebox(0,0)[cc]{$\big |\alpha _{l}\!\big >$}}
\put(40.77,38.23){\vector(1,1){0.2}}
\emline{39.70}{37.15}{7}{40.77}{38.23}{8}
\put(62.64,60.81){\vector(1,1){0.2}}
\emline{61.69}{59.86}{9}{62.64}{60.81}{10}
\end{picture}
\end{center}

\vspace{0.2cm}

\noindent
Explicitly, they are:
\beq
\big |\alpha _{l}\big >=
\left ( \begin{array}{c}
\tau (l,l+1) \\ \\ \hat \tau (l,l+1) \end{array}\right )\,,
\;\;\;\;\;\;\;\;
\big |\beta _{l}\big >=
\left ( \begin{array}{c}
\hat \tau (l+1,l) \\ \\ \tau (l+1,l) \end{array}\right )\,.
\label{q11}
\eeq
The normalization factor in eq.\,(\ref{r4}) is equal to
$\lambda _{l}=\mu \nu (\mu - \nu )^{-1}
\tau (l,l)\hat \tau (l,l)$.

\section{On the continuous time limit}

In this section we show that the $r$-matrix formula for
local $M$-operators (\ref{r6}) is a degenerate case of
eq.\,(\ref{r8}). As it was already mentioned at the end
of Sect.\,3, $\frac{\mu -\nu}{\mu \nu}$ plays the role of
lattice spacing of the discrete time variable $m$.
A naive continuous time limit would then be
$\nu \rightarrow \mu$ that means $q\rightarrow 1$ in the
$R$-matrix, so the first non-trivial term in its expansion
in $q-1$ is just the classical $r$-matrix. This agrees with
eq.\,(\ref{r6}). However, this limiting procedure would imply
$\lim _{q\rightarrow 1}\big |\check \beta _{l}\big >=
\big |\alpha _{l}\big >$ that is certainly wrong in general
(see e.g. the figure in the previous section). The correct
limiting transition to a continuous time coordinate needs
some clarification.

The limit $\nu \to \mu$ changes parameters of the space lattice
and the $L$-operator.
That is the reason why the naive limit does not work.
In the correct limiting procedure, the time lattice spacing
must approach zero independently of $\mu$ and $\nu$.

To achieve the goal, we introduce $v'$ -- another "copy" of
the chiral flow $v$ with the parameter $\nu '$, so
now we have a 3D lattice. Equations (\ref{b2})
are valid in each 2D section of this lattice of the form
$v' =\mbox{const}$,
$u=\mbox{const}$ or $v=\mbox{const}$.
For instance, in the latter case we have (cf. (\ref{b2})):
\begin{eqnarray}
&&\!\!(\nu '\!- \!\mu )\hat \tau (u,\!v,\!v')\tau (u\!+\!1,\! v,\!
v'+1)\!=\!  \nu ' \tau (u\!+\!1,\! v, \!v')
\hat \tau (u,\!v,\! v'\! +1) -\mu
\hat \tau (u\!+\!1,\! v,\! v')\tau (u,\!v, \!v'\! +\!1)\,,
\nonumber \\
&&\!\!(\nu '\!- \!\mu )
\tau (u,\!v,\!v')\hat \tau (u\!+\!1,\! v,\! v'\!+\!1)\!= \!
\nu ' \hat \tau (u\!+\!1,\! v, \!v')
\tau (u,\!v,\! v' \!+\!1) -\mu
\tau (u\!+\!1,\! v,\! v')\hat \tau (u,\!v,\! v'\! +\!1)\,.
\label{tb2}
\end{eqnarray}
Now we can tend $\nu ' \to \mu$ leaving $\nu$
unchanged.  Set
\beq
q' =\frac{\mu}{\nu '}=1+\varepsilon +O(\varepsilon^{2})\,,
\;\;\;\;\;\;\;\;
\varepsilon \to 0\,.
\label{t1}
\eeq
The small
parameter $\varepsilon$ has the meaning of lattice
spacing in the auxiliary direction
$m' =\frac{1}{2}(u-v' )$.

Discrete $M$-operators are defined up to multiplication
by a scalar function of $z$ independent of dynamical variables.
To pursue the continuous time limit, it is convenient to
normalize the $M$-operator in such a way that
${\cal M}_{l}(z)=I$ at $\varepsilon =0$. Then the next term
(of order $\varepsilon$)
yields a local $M$-operator of continuous time flow.
To find such an $M$-operator
at the $l$-th site of the 1D lattice
(diagonally embedded into the $u,v$-lattice as
$(l,l)$),
we should expand the discrete $M$-operator
${\cal M}_{B_{l}' \leftarrow C_{l}'}(z)$ which generates
the shift $(l-1, l, 1)\longrightarrow (l,l,0)$ on the
3D lattice with coordinates $(u,v,v')$.
This is illustrated by the figure

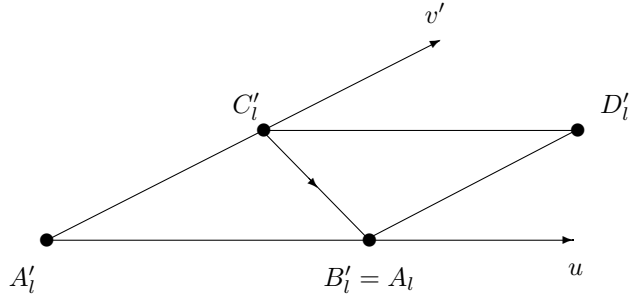
\begin{center}
\special{em:linewidth 0.4pt}
\unitlength 0.7mm
\linethickness{0.4pt}
\begin{picture}(127.50,66.25)(10.00,00.00)
\put(119.33,20.00){\vector(1,0){0.2}}
\emline{19.67}{20.00}{1}{119.33}{20.00}{2}
\emline{60.42}{40.83}{3}{120.00}{40.83}{4}
\emline{120.00}{40.83}{5}{80.00}{20.00}{6}
\emline{60.42}{40.83}{7}{80.83}{20.00}{8}
\put(70.86,30.16){\vector(1,-1){0.2}}
\emline{69.32}{31.77}{9}{70.86}{30.16}{10}
\put(19.58,20.00){\circle*{2.50}}
\put(60.83,40.83){\circle*{2.50}}
\put(120.42,40.83){\circle*{2.50}}
\put(80.83,20.00){\circle*{2.50}}
\put(15.00,12.50){\makebox(0,0)[cc]{$A_{l}'$}}
\put(80.83,12.50){\makebox(0,0)[cc]{$B_{l}' =A_l$}}
\put(57.50,45.42){\makebox(0,0)[cc]{$C_{l}'$}}
\put(127.50,45.42){\makebox(0,0)[cc]{$D_{l}'$}}
\put(120.00,14.50){\makebox(0,0)[cc]{$u$}}
\put(94.17,57.92){\vector(2,1){0.2}}
\emline{19.58}{20.00}{11}{94.17}{57.92}{12}
\put(93.33,63.25){\makebox(0,0)[cc]{$v'$}}
\end{picture}
\end{center}

\noindent
which shows the $u,v'$-section of the 3D lattice
spanned by $u,v,v'$.
Coordinates of the parallelogram vertices are:
$A_{l}'=(l-1,l,0)$, $B_{l}'=A_l =(l,l,0)$,
$C_{l}'=(l-1,l,1)$, $D_{l}'=(l,l,1)$.
The point $C_{l}'$ tends to the point
$B_{l}' =A_l$ as $\nu ' \to \mu $, so the parallelogram collapses
to the $u$-axis.
We have:
\beq
{\cal M}_{B_{l}' \leftarrow C_{l}'}(z)=I+\varepsilon M_{l}(z)+
O(\varepsilon ^{2})\,,
\label{t2}
\eeq
where
\beq
M_{l}(\mu z)=\frac{1}{z-z^{-1}}\left (
\begin{array}{ccc}
\displaystyle{\frac{1}{2}}
(z+z^{-1})\displaystyle{\frac{\tau (l-1,l)\hat \tau (l+1,l)}
{\tau (l,l)\hat \tau (l,l)}}&&
\displaystyle{\frac{\tau (l-1,l) \tau (l+1,l)}
{\tau (l,l)\hat \tau (l,l)}}\\&&\\
\displaystyle{\frac{\hat \tau (l-1,l) \hat \tau (l+1,l)}
{\tau (l,l)\hat \tau (l,l)}}&&
\displaystyle{\frac{1}{2}}
(z+z^{-1})\displaystyle{\frac{\hat \tau (l-1,l)\tau (l+1,l)}
{\tau (l,l)\hat \tau (l,l)}}\end{array}\right ).
\label{t3}
\eeq

By definition, the classical $r$-matrix is
\begin{eqnarray}
r(z)&=&\lim _{\varepsilon \to 0}\frac{R^{(+)}(z;q')-
(z-z^{-1})I\otimes I}{\varepsilon (z-z^{-1})}
\nonumber \\
&=&\frac{1}{2(z-z^{-1})}\left [ (z+z^{-1})I\otimes I+
2\sigma _{1}\otimes \sigma _{1}
+2\sigma _{2}\otimes \sigma _{2}
+(z+z^{-1})\sigma _{3}\otimes \sigma _{3}\right ].
\label{r}
\end{eqnarray}
Comparing with (\ref{t3}),
we obtain the $r$-matrix representation
\beq
M_l (\mu z)=
\frac{ \big < \beta _{l}
\big | r(z )\big |\alpha _{l-1} \big >}
{\big < \beta _{l}\big | \alpha _{l-1}\big >}
\label{t4}
\eeq
of the $M$-operator (see (\ref{r6})) with the trigonometric
classical $r$-matrix (\ref{r}).

\section{Conclusion}

The main result of this work is the $R$-matrix representation
(\ref{q9}), (\ref{q10}) of the local $L$-$M$ pair for the
classical lattice SG model with discrete space-time coordinates.
In our opinion, the very fact that the typical quantum $R$-matrix
naturally arises in a purely classical problem is
important and interesting by itself. As a by-product, we
have shown that components of the vectors $\big |\alpha \big >$,
$\big |\beta \big >$ (representing the $L$-operator at the
degeneracy point, see (\ref{r4})) are $\tau$-functions.

Let us recall that the quantum Yang-Baxter equation already
appeared in connection with purely classical problems,
though in a different
context \cite{Sklyanin85},\,\cite{WX}. However, the class
of solutions relevant to classical
problems is most likely very far from
$R$-matrices of known quantum integrable models.
In our construction, the role of the quantum
Yang-Baxter equation remains obscure; instead,
the most popular 4$\times$4 trigonometric quantum $R$-matrix is
shown to take part in the zero curvature representation of the
classical discrete SG model. We believe that a conceptual
explanation of this phenomenon nevertheless relies on the
Yang-Baxter equation.

We should stress that the "quantum deformation parameter" $q$
of the trigonometric $R$-matrix in our context
seems to have nothing to do with any kind of quantization.
Indeed, it is related to the mass parameter and the
lattice spacing of the classical model. This fact suggests
to ask for a non-standard hidden $R$-matrix
structure of the
model, which might survive and show up on the quantum
level, too. Another interesting question is to
find a discrete time analogue of the non-local generating
function (\ref{r3}) of $M$-operators.

At last, let us point out that the $R$-matrix representation
can be extended to $M$-operators of
a more general model -- partially
anisotropic classical Heisenberg spin chain in discrete time.
This system includes the discrete KdV and some other
equations as particular cases.
The results will be reported elsewhere.

\section*{Acknowledgements}

I thank S.Kharchev and P.Wiegmann for permanent
interest to this work, very helpful discussions and critical
remarks. Discussions with O.Lipan, I.Krichever and A.Volkov
are also gratefully acknowledged.
This work was supported in part by RFBR
grant 97-02-19085.

\end{document}